\documentclass[aps,prl,reprint]{revtex4-1}
\usepackage{graphicx}
\usepackage{times}
\usepackage{bm}
\usepackage{amsmath,amssymb}
\usepackage{feynmf}

\begin{document}
\title{Proton size from precision experiments on hydrogen and muonic hydrogen atoms}
\author{D. Solovyev}
\email[E-mail:]{d.solovyev@spbu.ru}
\affiliation{ 
Department of Physics, St. Petersburg State University, Petrodvorets, Oulianovskaya 1, 198504, St. Petersburg, Russia
}
\author{T. Zalialiutdinov}
\affiliation{ 
Department of Physics, St. Petersburg State University, Petrodvorets, Oulianovskaya 1, 198504, St. Petersburg, Russia
}
\author{A. Anikin}
\affiliation{ 
Department of Physics, St. Petersburg State University, Petrodvorets, Oulianovskaya 1, 198504, St. Petersburg, Russia
}

\begin{abstract}
The "proton radius puzzle" was recently solved by reducing the four-standard deviation discrepancy between the results for electronic hydrogen ($H$) and muonic hydrogen ($\mu H$) atoms to $3.3$ value. The value of the root-mean-square radius of the proton ($r_p$), extracted from experiments on measuring the one-photon $2s-4p$ transition and the Lamb shift in hydrogen, is now $0.8335(95)$ fm, that is in good agreement with the muonic hydrogen experiments, $0.84087(39)$ fm. Even so, these values deviate significantly from the CODATA value, which is determined as the average using the results for various spectral lines including two-photon transitions in the hydrogen atom. The solution of the proton radius puzzle was realized by taking into account the influence of interference effect in one-photon scattering processes. The importance of interfering effects in atomic frequencies measurements gives an impetus to the study of experiments based on two-photon spectroscopy with the suchlike thoroughness. It is shown here that the effect of interfering pathways for two-photon $2s-nd$ transitions in a hydrogen atom is also significant in determining the proton charge  radius and Rydberg constant. 
\end{abstract}

\maketitle


The partway solved problem of the "proton radius puzzle" is still of interest to modern atomic physics. The problem arose with the publication of work \cite{Pohl}, where it was first reported that the value of the root-mean-square (rms) radius of the proton ($r_p$) is $r_p = 0.84184(67)$ fm. This rms value, extracted from the experiments with muonic hydrogen, deviates more than a four standard deviation from the proton charge radius accepted by the CODATA \cite{Mohr-2016,Mohr-2016-RMP}, $r_p=0.8751(61)$ fm. The latter is obtained as the average value of $r_p$ extracted from spectroscopic measurements of various transitions in the hydrogen atom, and also includes the result found in \cite{Bernauer} for electron-proton ($e-p$) elastic scattering data. The reasons for the controversy remained unknown for a decade before the recent study of the one-photon $2s-4p$ transition in a hydrogen atom \cite{H-exp}. Acting as a reference point, such an experiment served as a driving force for obtaining similar values in measurements of the Lamb shift in a hydrogen \cite{LambShift-H} and electron-proton scattering experiments \cite{e-p_scatt}. Although the $r_p$ values obtained from measurements of the one-photon $2s-4p$ transition and the Lamb shift in the hydrogen atom are in excellent agreement with each other, the discrepancy on the level of $3.3$ standard deviation with muonic hydrogen is in question.

The obtaining of the proton charge radius involves the joint calculation of the Rydberg constant in the hydrogen atom, that expressed by the dependence of energy of the bound states on these parameters:
\begin{eqnarray}
\label{1}
E_{nlj} = R_\infty\left(-\frac{1}{n^2}+f_{nlj}\left(\alpha,\frac{m_e}{m_p},r_p,\dots\right)\right),
\end{eqnarray}
where $n$, $l$ and $j$ are the principal, orbital and total angular momentum quantum numbers, respectively. $R_\infty$ = $m_e\alpha^2c/2h$ is the Rydberg constant ($c$ is the speed of light, $h$ is the Planck’s constant and $\alpha$ is the fine structure constant), $m_e$ and $m_p$ represent the electron and proton masses. The function $f_{nlj}$ denotes all the possible corrections arising within the relativistic QED theory, see \cite{Mohr-2016,Mohr-2016-RMP}. 

To determine the Rydberg constant and proton radius the theoretical results are compared with the corresponding experimental data: $E_{nlj} - E_{n'l'j'} = \Delta E_{nlj - n'l'j'}^{\rm exp}$. Assuming that there are only two unknown constants, $r_p$ and $R_\infty$, the set of equations for two independent transitions should be constructed from the equality above. As a rule, one of them corresponds to the most precisely determined $1s-2s$ transition in a hydrogen atom \cite{Mat}, the $1s-3s$ transition frequency \cite{arnoult-1s-3s,Fleurbaey-1s-3s} is also appropriate for this purpose.

Prior to \cite{H-exp} various relativistic QED corrections were included in the formula (\ref{1}) only. A nonresonant effect, called quantum interference effect (QIE), was used to determine the new values of $r_p = 0.8335(95)$ fm and $R_\infty = 10973731.568076(96)$ m$^{-1}$ in \cite{H-exp}. The presence of nonresonant (NR) effects has been shown by F. Low in \cite{Low}. Then, later, in \cite{LKG, LGL}, the importance of nonresonant corrections for the Lamb shift measurements and the process of radiative electron capture in highly charged hydrogen-like ions was demonstrated. In further NR corrections were evaluated for the hydrogen atom in \cite{LSPS,can-LSPS,PRA-LSPS,PRL-LSSP,LSSCK-2009}, whereas the most essential NR contribution was found in \cite{Jent-Mohr}. It is the nonresonant correction including the fine structure splitting of atomic levels \cite{PRA-LSPS,Jent-Mohr} represents a special interest in further investigations, see, for example, \cite{HH-2010,HH-2011,Sansonetti,Brown-2013,MHH-2015,Amaro-2015,Amaro-mH-2015}. Recently a theoretical analysis of the experiment \cite{H-exp} was presented in \cite{S-2020-importance}, where it was shown that QIE effects should be carefully studied with each specific spectroscopic measurement.


Measurements of the Lamb shift in muonic hydrogen were re-implemented in \cite{Antognini-2013,Antog-2013}, where the value of the proton charge radius $r_p = 0.84087(39)$ fm was reported. This value is insensitive to the QIE, see \cite{Amaro-mH-2015} and has been refined with a mixing effect of the $2^{2F+1}p$ states ($F$ is the total momentum), see \cite{Pachucki}. For this purpose the two hyperfine splitted transition frequencies $2^1s_{1/2}\rightarrow 2^3p_{3/2}$, $\nu_s$ (singlet), and $2^3s_{1/2}\rightarrow 2^5p_{3/2}$, $\nu_t$ (triplet), were measured. Including the hyperfine splitting (but still without the mixing effect), one can find \cite{Antog-2013}
\begin{eqnarray}
\label{2}
h\nu_s = E_L +\Delta_{\rm fs}+\frac{3}{4}\Delta_{\rm hfs}^{2S_{1/2}}-\frac{5}{8}\Delta_{\rm hfs}^{2P_{3/2}}
\\
\nonumber
h\nu_t = E_L +\Delta_{\rm fs} -\frac{1}{4}\Delta_{\rm hfs}^{2S_{1/2}}+ \frac{3}{8} \Delta_{\rm hfs}^{2P_{3/2}}.
\end{eqnarray}

To obtain the proton charge radius, it is necessary compare the theoretical results for the Lamb shift $E_L = 206.0668 - 5.2275 r_p^2$ meV and for the hyperfine splitting of the $2s$ state $22.9843 - 0.1621 r_Z$ ($r_Z$ is the proton Zemach radius) with the experimentally measured frequencies $\nu_s = 54611.16(1.05)$ GHz and $\nu_t = 49881.35(65)$ GHz \cite{Antognini-2013,Antog-2013}. Then, using the values of the theoretical predictions for fine splitting, $\Delta_{\rm fs}=8.352082$ meV, and hyperfine splitting of the $2p_{3/2}$ state, $\Delta_{\rm hfs}^{2P_{3/2}}=3.392588$ meV, including the mixing shift $\delta = 0.14456$ meV \cite{Pachucki}, see also \cite{Martynenko-2008}, as it was given in \cite{Antognini-2013,Antog-2013} (i.e. inserting $\delta$ only in equation for the singlet line, Eq. (\ref{2})), one can find the same result: $r_p=0.84087$ fm and $r_Z = 1.082$ fm.
However, this treatment can lead to confusion in combination with the corresponding scheme of transitions in muonic hydrogen \cite{Antognini-2013,Antog-2013}. A more accurate analysis was carried out in \cite{Karshenboim}, where the same value of the proton charge radius was found.

As discussed above, determining the proton radius from spectroscopic measurements requires a comprehensive analysis of the QI effects. For the singlet transition $2^1s_{1/2}\rightarrow 2^3p_{3/2}$ there is interference between the states $2^3p_{3/2}$ and $2^3p_{1/2}$, and for the triplet transition $2^3s_{1/2}\rightarrow 2^5p_{3/2}$, the interference occurs taking into account the states $2^1p_{1/2}$, $2^3p_{1/2}$, and $2^3p_{3/2}$. Analysis in \cite{Amaro-mH-2015} has shown that the effect of quantum interference is not important when measuring the $\nu_s$ and $\nu_t$ frequencies in muonic hydrogen. Applying the results obtained in \cite{S-2020-importance}, the maximum NR correction values are $\delta_{{\rm NR}}^{s}= -7.40\times 10^{7}$ Hz (or $-3.06\times 10^{-4}$ meV) and $\delta_{{\rm NR}}^{t}=-5.26\times 10^{7}$ Hz (or $-2.18\times 10^{-4}$ meV) for singlet and triplet transitions, respectively. Substitution of these corrections to the left side of Eqs. (\ref{2}) makes  this conclusion reliable, and the values of the radii are predominantly stable.


There is still the problem of calculating the proton charge radius taking into account transitions to higher excited states in the hydrogen atom, the measurements of which are carried out using two-photon spectroscopy. The first attempts to calculate NR corrections to them were made in \cite{PRA-LSPS,Jent-Mohr}, where the negligibly small contributions for the $1s-2s$ transition frequency with respect to the current level of experimental accuracy were found. Theoretical analysis of such experiments is complicated by the presence of an external electric field acting on the excited atom with a time delay, see \cite{LSSPS-2003,PRL-LSSP,LSSCK-2009}. Reference can also be made to the result of \cite{PRL-LSSP}, where the nonresonant correction $0.17$ MHz for the Lyman-$\alpha$ spectral line was found taking into account the hyperfine splitting, while the uncertainty of the frequency measurement is about $6$ MHz. Since the effect of quantum interference is sensitive to the hyperfine structure of levels, below we analyze the two-photon absorption transitions similarly to \cite{PRA-LSPS,Jent-Mohr,HH-2010,HH-2011,Sansonetti,Brown-2013,MHH-2015,Amaro-2015,Amaro-mH-2015,S-2020-importance}.

Measurements of the $2s-ns/nd$ transition frequencies based on two-photon spectroscopy can be divided into two different types of experiments. Namely, one of them corresponds to the detection of the excited $nd$ fraction of atoms observed via its fluorescence (i.e.
decay to the 2p state) \cite{Weitz} or the decrease in metastable 2s signal \cite{Weitz-1, Nez}. Nonresonant corrections to experiments of this type were recently evaluated in \cite{Anikin}. Another type of two-photon spectroscopy experiments can be found in  \cite{deB-0,Schwob,deB-1,deB-2}, which we will focus on.

The theoretical description of experiments \cite{deB-0,Schwob,deB-1,deB-2} can be attributed to the process of multi-photon scattering, when measurements of $2s - nd$ frequencies ($n$ is the principal quantum number equal to 4, 6, 8 or 12) correspond to the observation of a two-photon absorption profile. Considering the photon emission process as an indicator for recording absorption, it is sufficient to describe only the absorption profile. This technique has been used to obtain the two-photon excitation rate in \cite{deB-2} (see section 3).

Within the framework of the $S$-matrix formalism, the amplitude of two-photon absorption after successive and standard calculations is
\begin{eqnarray}
\label{4}
&&U_{ai}^{\rm abs} = e^2\frac{2\pi\sqrt{\omega_1\omega_2}}{E_i+\omega_1+\omega_2-E_a(1-i0)}\times
\\
\nonumber
&&\left[\sum\limits_k\frac{\langle a| \vec{e}_1\vec{r}|k\rangle\langle k|\vec{e}_2\vec{r}|i\rangle}{E_a-\omega_1-E_k(1-i0)} + \sum\limits_k\frac{\langle a| \vec{e}_2\vec{r}|k\rangle\langle k|\vec{e}_1\vec{r}|i\rangle}{E_i+\omega_1-E_k(1-i0)}\right].
\end{eqnarray}
Here $\vec{e}_j$ ($j=1,2$) denote the polarization vectors of the absorbed photons and $\omega_j$ their frequencies, $E_i$ and $E_a$ are the energies of the initial and excited states, and the sum runs the entire spectrum. The second term in Eq. (\ref{4}) corresponds to the permutation of photons.

Omitting the intermediate calculations for brevity, which include integration over the angles and summation over the projections, each term in Eq. (\ref{4}) can be reduced to
\begin{eqnarray}
\label{5}
&&\sum\limits_{k}\frac{\langle a| \vec{e}_2\vec{r}|k\rangle\langle k|\vec{e}_1\vec{r}|i\rangle}{E_i+\omega_1-E_k(1-i0)} = (-1)^{l_k+l_i+j_a+2j_k+F_a+j_i+F_k} \times
\nonumber
\\
\nonumber
&&\sqrt{(2l_i+1)(2l_k+1)(2j_i+1)(2j_a+1)} \times 
\\
&&(2j_k+1)\sqrt{(2F_k+1)(2F_i+1)}C^{l_a\,0}_{l_k\,0\,\,\, 1\,0}C^{l_k\,0}_{l_i\,0\,\,\, 1\,0}\times 
\\
\nonumber
&& \left\{ 
\begin{array}{ccc}
l_k & s & j_k\\
j_a & 1 & l_a
\end{array}\right\}
\left\{
\begin{array}{ccc}
l_i & s & j_i\\
j_k & 1 & l_k
\end{array}\right\}
\left\{
\begin{array}{ccc}
j_k & I & F_k\\
F_a & 1 & j_a
\end{array}\right\}
\left\{
\begin{array}{ccc}
j_i & I & F_i\\
F_k & 1 & j_k
\end{array}\right\} \times
\\
\nonumber
&&\sum\limits_{q_1,q_2}(-1)^{q_1+q_2}C^{F_a\,M_a}_{F_k\,M_k\,\,\, 1\,-q_1}C^{F_k\,M_k}_{F_i\,M_i\,\,\, 1\,-q_2}e_{1_{q_1}}e_{2_{q_2}}g_{l_k}(E_i+\omega).
\end{eqnarray}
Here, the summation over $k$ in the left side of the expression means all the necessary quantum numbers not included in the right side, $e_{1(2)_q}$ represents the spherical component of the polarization vector, coefficients $\left\{ \begin{array}{ccc}
j_1 & j_2 & j_{12}\\
j_3 & j & j_{23}
\end{array}\right\}$ are the $6j$ Wigner coefficients, $C^{l\,m}_{l_1\,m_1\,\,\, l_2\,m_2}$ is the Clebsch–Gordan coefficient, $F$ represents the total momentum with the projection denoted as $M$, $j$ is the total angular momentum and $l$ is the orbital momentum. The function $g_{l}(E)$ is the result of the radial integration $g_{l_k}(E_i+\omega) = \int\limits_0^\infty\int\limits_0^\infty dr_1 dr_2 R_{n_a l_a}(r_1)r^3_1 g_{l_k}(E_i+\omega;r_1, r_2)r^3_2 R_{n_i l_i}(r_2)$, where $g_{l_k}(E_i+\omega;r_1, r_2)$ is the radial part of the Green function, see, for example, \cite{Andr} and references therein. 
The absorption rate can be obtained with the use of relation $dW_{ai}^{\rm abs} = \frac{d^3k_1}{(2\pi)^3}\frac{d^3k_2}{(2\pi)^3}\left|U_{ai}^{\rm abs}\right|^2$,
where $d^3k_j/(2\pi)^3$ represents the phase volume of corresponding photon.

Further, one can use the resonance approximation with $\omega_1=\omega_2 \equiv \omega =  (E_a-E_i)/2$ \cite{Andr,ZSLP-report}. As a result of regularization by QED methods, see \cite{Low,Andr}, of the divergent denominator in the common factor of Eq. (\ref{4}), the Lamb shift and level width of the excited state $a$ arise in the energy denominator as the real and imaginary parts, respectively. The appearance of the imaginary part leads to the formation of the absorption line profile. When the absorption profile is set out, the frequency in all expressions can be replaced by its resonant value without loss of accuracy, for more details see \cite{Andr}. Such an approximation is justified by the fact that the corresponding nonresonant corrections go beyond the accuracy of experiments \cite{deB-0,Schwob,deB-1,deB-2}.

The most significant nonresonant contribution (QIE) occurs taking into account the fine structure of the excited levels, when the state $b$ with the same total momentum but with a different total angular momentum (states $nd_{3/2}$ and $nd_{5/2}$ for example) is added in the amplitude (\ref{4}), see \cite{S-2020-importance,Anikin}. Then the absorption rate can be written as
\begin{eqnarray}
\label{6}
dW_{ai}^{\rm abs} \sim \frac{C_a d\omega}{(2\omega-\omega_0)^2+\frac{1}{4}\Gamma^2_a} + \nonumber
\\
\frac{C_b d\omega}{(2\omega-\omega_0-\Delta_{fs})^2+\frac{1}{4}\Gamma^2_b}+
\\
\nonumber
C_{ab}\frac{2(2\omega-\omega_0)}{2\omega-\omega_0-\Delta_{fs}}\frac{d\omega}{(2\omega-\omega_0)^2+\frac{1}{4}\Gamma^2_a}.
\end{eqnarray}
Here $\Gamma_{a(b)}$ is the natural level width of the corresponding state, $\Delta_{fs} = E_{nd_{3/2}}(F=2) - E_{nd_{5/2}}(F=2)$ denotes the fine splitting energy including the hyperfine splitting and $\omega_0 = E_a-E_i$. The coefficients $C_a$, $C_b$ and $C_{ab}$ should be calculated in accordance with Eqs. (\ref{4}), (\ref{5}). The first term here represents the absorption line profile for the transition under study, the second corresponds to absorption along the second pathway and, finally, the third term represents the contribution of interference between them.

One evident way to define the resonant frequency $\omega_{\rm res}$ corresponds to the search of maximum for the line profile Eq. (\ref{7}), see the discussion in \cite{S-2020-importance}. Then the leading order NR corrections can be found from the extremum condition $dW_{ai}^{\rm abs}/d\omega=0$ with $\omega_{\rm res} = \omega_0+\delta_{\rm NR}$. The result is
\begin{eqnarray}
\label{7}
\delta_{\rm NR} = -\frac{C_{ab}\Gamma_{nd}^2}{4C_a \Delta_{fs}} + 
O\left(\frac{\Gamma_{nd^4}}{\Delta_{fs}^3}\right)
.
\end{eqnarray}
It can also be emphasized that the summation over polarizations and subsequent integration over photon directions are not necessary in Eq. (\ref{6}). In the final expression for the cross-section their combination gives a common factor, which is eliminated in the nonresonant correction. This is the result of an approximation leading to the independence of absorption process from radiation.

The contribution Eq. (\ref{7}) was called the quantum interference effect. The next order correction can be obtained from Eq. (\ref{6}), see \cite{PRA-LSPS}, but it does not exceed a few hertz, and we exclude its further consideration. Some results for interfering transitions $2s_{j_i=1/2}(F_i=1) \rightarrow nd_{j_a =3/2}(F_a=2)$ and $2s_{j_i=1/2}(F_i=1) \rightarrow nd_{j_a=5/2}(F_a=2)$ in hydrogen are listed in Table~\ref{tab:1} in approximation of equal widths,
 where the uncertainties of the corresponding two-photon transitions are also indicated \cite{Mohr-2016-RMP}.
\begin{center}
\begin{table}
\caption{Nonresonant correction corresponding to Eq. (\ref{7}). The first column shows the excited state $a$, the second and third columns contain the used values of energy splitting, $\Delta_{fs}$ \cite{HH-tab}, and the level width, $\Gamma_{nd}$, respectively. The values of $\delta_{\rm NR}$ are collected in the fifth column. In the last column the values of uncertainties for the $2s-nd_{5/2}$ transitions are taken from \cite{Mohr-2016}. All numbers are given in Hz.}
\begin{tabular}{l | c | c | c | c | c }
\hline
\hline
state & $\Delta_{fs}$ in Hz & $\Gamma_{nd}$ in Hz & $\delta_{\rm NR}$ in Hz & Unc. in Hz\\
\hline

$4d$ & $4.557026\times 10^8$ & $4.40503\times 10^6$ & $-8691.82$ & $24.\times 10^3$\\

$6d$ & $1.350231\times 10^8$ & $1.33682\times 10^6$ & $-2701.67$ & $10.\times 10^3$\\

$8d$ & $5.69628\times 10^7$ & $5.72382\times 10^5$ & $-1174.02$ & $6.4\times 10^3$\\

$12d$ & $1.68779\times 10^7$ & $1.72261\times 10^5$ & $-358.88$ & $7.0\times 10^3$\\

\hline
\hline
\end{tabular}
\label{tab:1}
\end{table}
\end{center}


Finally, the NR corrections can be considered in connection with the problem of determining the Rydberg constant and the proton charge radius. For this purpose, we use Eq. (\ref{1}) and a pair of transitions: $1s-2s/3s$ combined with $2s-nd_{3/2(5/2)}$. The results are collected in Table~\ref{tab:2}. Our calculations are divided into three parts. The first part corresponds to the estimation of the values $R_\infty$ and $r_p$ according to the data given in \cite{Mohr-2016-RMP}. The second part presents the values of the Rydberg constant and the proton charge radius, denoted by $R_\infty^{\rm HH}$ and $r_p^{\rm HH}$, matching the values and analysis given in Table VII from \cite{HH-tab}. Finally, the third part is obtained using the data of \cite{HH-tab} combined with the nonresonant correction from Table~\ref{tab:1}.
\begin{widetext}
\begin{center}
\begin{table}
\caption{Rydberg constant, $R_\infty$, and proton charge radius, $r_p$. The pair of transitions used to determine $R_\infty$ and $r_p$ is shown in the first column. The values of $R_\infty$, $r_p$ are given in the second and sixth (CODATA), third and seventh \cite{HH-tab}, fourth and eighth columns contain the values evaluated with \cite{HH-tab} and NR corrections. The notations $a=\frac{2}{5}(2s-8d_{3/2})+\frac{3}{5}(2s-8d_{5/2})$, $b=\frac{2}{5}(2s-12d_{3/2})+\frac{3}{5}(2s-12d_{5/2})$, $c=2s-4d_{5/2} -\frac{1}{4} (1s-2s)$, $d=2s-6d_{5/2} -\frac{1}{4} (1s-3s)$ are introduced. The uncertainties of found values, estimated using the experimental errors of the CODATA values, are given in fifth and ninth columns for the Rydberg constant and proton charge radius, respectively.}
\begin{tabular}{l | c | c | c || c ||c | c | c || r}
\hline
\hline
state & $R_\infty$, m$^{-1}$ & $R_\infty^{\rm HH}$, m$^{-1}$ & $R_\infty^{{\rm HH + NR}}$, m$^{-1}$ & $\delta R_\infty$, m$^{-1}$ & $r_p$, fm & $r_p^{\rm HH}$, fm & $r_p^{{\rm HH + NR}}$, fm & $\delta r_p$, fm \\
\hline

$1s-2s, 2s-8d_{3/2}$ & $10973731.568548$ & $10973731.568152$ & $10973731.568121$ & $0.000096$ & $0.87904$ & $0.84123$ & $0.8382$ & $0.0094$ \\
$1s-3s, 2s-8d_{3/2}$ & $10973731.568528$ & $10973731.568105$ & $10973731.568072$ & $0.000094$ & $0.87503$ & $0.83112$ & $0.8267$ & $0.0087$\\
\hline

$1s-2s, 2s-8d_{5/2}$ & $10973731.568681$ & $10973731.568153$ & $10973731.568184$ & $0.000072$ & $0.89133$ & $0.84135$ & $0.8443$ & $0.0070$\\
$1s-3s, 2s-8d_{5/2}$ & $10973731.568670$ & $10973731.568106$ & $10973731.568139$ & $0.000070$ & $0.88923$ & $0.83127$ & $0.8348$ & $0.0060$\\
\hline
\hline
$1s-2s, a$ & $10973731.568429$ & $10973731.567954$ & $10973731.567960$ & $0.000082$ & $0.86782$ & $0.82167$ & $0.8223$ & $0.0081$\\
$1s-3s, a$ & $10973731.568401$ & $10973731.567893$ & $10973731.567900$ & $0.000081$ & $0.86203$ & $0.80827$ & $0.8089$ & $0.0081$\\
\hline
\hline

$1s-2s, 2s-12d_{3/2}$ & $10973731.568297$ & $10973731.568152$ & $10973731.568144$ & $0.000104$ & $0.85529$ & $0.84126$ & $0.8404$ & $0.0101$\\
$1s-3s, 2s-12d_{3/2}$ & $10973731.568263$ & $10973731.568109$ & $10973731.568099$ & $0.000104$ & $0.84775$ & $0.83150$ & $0.8305$ & $0.0105$\\
\hline

$1s-2s, 2s-12d_{5/2}$ & $10973731.568392$ & $10973731.568151$ & $10973731.568160$ & $0.000075$ & $0.86433$ & $0.84115$ & $0.8420$ & $0.0073$\\
$1s-3s, 2s-12d_{5/2}$ & $10973731.568364$ & $10973731.568107$ & $10973731.568117$ & $0.000074$ & $0.85820$ & $0.83138$ & $0.8324$ & $0.0070$\\
\hline
\hline

$1s-2s, b$ & $10973731.568410$ & $10973731.568208$ & $10973731.568226$ & $0.000087$ & $0.86608$ & $0.84668$ & $0.8484$ & $0.0084$\\
$1s-3s, b$ & $10973731.568383$ & $10973731.568168$ & $10973731.568187$ & $0.000086$ & $0.86022$ & $0.83778$ & $0.8398$ & $0.0084$\\
\hline
\hline

$1s-2s, c$ & $10973731.569110$ & $10973731.568138$ & $10973731.568058$ & $0.000705$ & $0.93003$ & $0.83984$ & $0.8319$ & $0.0696$\\
$1s-3s, c$ & $10973731.569074$ & $10973731.568133$ & $10973731.568113$ & $0.000654$ & $0.92853$ & $0.83405$ & $0.8319$ & $0.0696$\\
\hline

$1s-2s, d$ & $10973731.568308$ & $10973731.568153$ & $10973731.568062$ & $0.000246$ & $0.85628$ & $0.84130$ & $0.8324$ & $0.0243$\\
$1s-3s, d$ & $10973731.568345$ & $10973731.568201$ & $10973731.568117$ & $0.000226$ & $0.85628$ & $0.84130$ & $0.8324$ & $0.0243$\\
\hline



\hline
\hline
rms(1s-2s) & $10973731.568522$ & $10973731.568133$ & $10973731.568114$ & $0.000104$ & $0.87658$ & $0.83934$ & $0.8375$ & $0.0102$\\
rms(1s-3s) & $10973731.568503$ & $10973731.568103$ & $10973731.568093$ & $0.000097$ & $0.87250$ & $0.83088$ & $0.8297$ & $0.0102$\\
rms & $10973731.568513$ & $10973731.568118$ & $10973731.568103$ & $0.000068$ & $0.87454$ & $0.83512$ & $0.8336$ & $0.0070$\\
\hline
\hline
\end{tabular}
\label{tab:2}
\end{table}
\end{center}
\end{widetext}


Numerical results for the nonresonant correction Eq. (\ref{7}) are given in Table~\ref{tab:1}. Its role in determining the Rydberg constant and the proton charge radius can be found in Table~\ref{tab:2} for specific transitions. To verify the accuracy of our results, we calculated the root mean square values of $R_\infty$ and $r_p$ using CODATA \cite{Mohr-2016}. The obtained results are in complete agreement with the recommended $R_\infty=10973731.568508(65)$ m$^{-1}$ and $r_p=0.879(11)$ fm, see Table~\ref{tab:2}.  However, the analysis given in \cite{HH-tab} reveals the need for experimental resolution of hyperfine splitting in measurements of the type \cite{deB-0,deB-1,deB-2}. The rms values for the Rydberg constant and the proton charge radius based on the re-analyzed data in \cite{HH-tab} are $10973731.568118$ m$^{-1}$ and $0.83512$ fm, respectively.  Consolidating the \cite{HH-tab} results with the inclusion of nonresonant effects in the analysis the obtained rms values are $R_\infty=10973731.568103$ $m^{-1}$ and $r_p=0.83364$ fm. The latter is in perfect agreement with the result of \cite{H-exp}.

The main conclusion that follows from the analysis presented in this work consists in that nonresonant corrections Eq. (\ref{7}) for experiments such as \cite{deB-0,Schwob,deB-1,deB-2} can reach a level of several kHz, see Table~\ref{tab:1}. These corrections are an order of magnitude higher than the corresponding values found in \cite{Anikin} for experimental technique based on the fluorescence process \cite{Weitz,Weitz-1, Nez}. Thus, the interference effects described in this work must be taken into account when measuring the frequency in the experiments \cite{deB-0,Schwob,deB-1,deB-2}.

This work was supported by the Russian Science Foundation under grant No. 20-72-00003.
\bibliography{mybibfile} 

\begin{thebibliography}{44}%
\makeatletter
\providecommand \@ifxundefined [1]{%
 \@ifx{#1\undefined}
}%
\providecommand \@ifnum [1]{%
 \ifnum #1\expandafter \@firstoftwo
 \else \expandafter \@secondoftwo
 \fi
}%
\providecommand \@ifx [1]{%
 \ifx #1\expandafter \@firstoftwo
 \else \expandafter \@secondoftwo
 \fi
}%
\providecommand \natexlab [1]{#1}%
\providecommand \enquote  [1]{``#1''}%
\providecommand \bibnamefont  [1]{#1}%
\providecommand \bibfnamefont [1]{#1}%
\providecommand \citenamefont [1]{#1}%
\providecommand \href@noop [0]{\@secondoftwo}%
\providecommand \href [0]{\begingroup \@sanitize@url \@href}%
\providecommand \@href[1]{\@@startlink{#1}\@@href}%
\providecommand \@@href[1]{\endgroup#1\@@endlink}%
\providecommand \@sanitize@url [0]{\catcode `\\12\catcode `\$12\catcode
  `\&12\catcode `\#12\catcode `\^12\catcode `\_12\catcode `\%12\relax}%
\providecommand \@@startlink[1]{}%
\providecommand \@@endlink[0]{}%
\providecommand \url  [0]{\begingroup\@sanitize@url \@url }%
\providecommand \@url [1]{\endgroup\@href {#1}{\urlprefix }}%
\providecommand \urlprefix  [0]{URL }%
\providecommand \Eprint [0]{\href }%
\providecommand \doibase [0]{http://dx.doi.org/}%
\providecommand \selectlanguage [0]{\@gobble}%
\providecommand \bibinfo  [0]{\@secondoftwo}%
\providecommand \bibfield  [0]{\@secondoftwo}%
\providecommand \translation [1]{[#1]}%
\providecommand \BibitemOpen [0]{}%
\providecommand \bibitemStop [0]{}%
\providecommand \bibitemNoStop [0]{.\EOS\space}%
\providecommand \EOS [0]{\spacefactor3000\relax}%
\providecommand \BibitemShut  [1]{\csname bibitem#1\endcsname}%
\let\auto@bib@innerbib\@empty
\bibitem [{\citenamefont {Pohl}\ and\ \citenamefont {et.al.}(2010)}]{Pohl}%
  \BibitemOpen
  \bibfield  {author} {\bibinfo {author} {\bibfnamefont {R.}~\bibnamefont
  {Pohl}}\ and\ \bibinfo {author} {\bibnamefont {et.al.}},\ }\href {\doibase
  10.1038/nature09250} {\bibfield  {journal} {\bibinfo  {journal} {Nature}\
  }\textbf {\bibinfo {volume} {466}},\ \bibinfo {pages} {213} (\bibinfo {year}
  {2010})}\BibitemShut {NoStop}%
\bibitem [{\citenamefont {Mohr}\ \emph
  {et~al.}(2016{\natexlab{a}})\citenamefont {Mohr}, \citenamefont {Newell},\
  and\ \citenamefont {Taylor}}]{Mohr-2016}%
  \BibitemOpen
  \bibfield  {author} {\bibinfo {author} {\bibfnamefont {P.~J.}\ \bibnamefont
  {Mohr}}, \bibinfo {author} {\bibfnamefont {D.~B.}\ \bibnamefont {Newell}}, \
  and\ \bibinfo {author} {\bibfnamefont {B.~N.}\ \bibnamefont {Taylor}},\
  }\href {\doibase 10.1063/1.4954402} {\bibfield  {journal} {\bibinfo
  {journal} {J. Phys. Chem. Ref. Data}\ }\textbf {\bibinfo {volume} {45}},\
  \bibinfo {pages} {043102} (\bibinfo {year} {2016}{\natexlab{a}})}\BibitemShut
  {NoStop}%
\bibitem [{\citenamefont {Mohr}\ \emph
  {et~al.}(2016{\natexlab{b}})\citenamefont {Mohr}, \citenamefont {Newell},\
  and\ \citenamefont {Taylor}}]{Mohr-2016-RMP}%
  \BibitemOpen
  \bibfield  {author} {\bibinfo {author} {\bibfnamefont {P.~J.}\ \bibnamefont
  {Mohr}}, \bibinfo {author} {\bibfnamefont {D.~B.}\ \bibnamefont {Newell}}, \
  and\ \bibinfo {author} {\bibfnamefont {B.~N.}\ \bibnamefont {Taylor}},\
  }\href {\doibase 10.1103/RevModPhys.88.035009} {\bibfield  {journal}
  {\bibinfo  {journal} {Rev. Mod. Phys.}\ }\textbf {\bibinfo {volume} {88}},\
  \bibinfo {pages} {035009} (\bibinfo {year} {2016}{\natexlab{b}})}\BibitemShut
  {NoStop}%
\bibitem [{\citenamefont {Bernauer}\ \emph {et~al.}(2014)\citenamefont
  {Bernauer}, \citenamefont {Distler}, \citenamefont {Friedrich}, \citenamefont
  {Walcher}, \citenamefont {Achenbach}, \citenamefont {Ayerbe~Gayoso},
  \citenamefont {B\"ohm}, \citenamefont {Bosnar}, \citenamefont {Debenjak},
  \citenamefont {Doria}, \citenamefont {Esser}, \citenamefont {Fonvieille},
  \citenamefont {GomezRodriguezdelaPaz}, \citenamefont {Friedrich},
  \citenamefont {Makek}, \citenamefont {Merkel}, \citenamefont {Middleton},
  \citenamefont {M\"uller}, \citenamefont {Nungesser}, \citenamefont
  {Pochodzalla}, \citenamefont {Potokar}, \citenamefont {S\'anchez~Majos},
  \citenamefont {Schlimme}, \citenamefont {\ifmmode~\check{S}\else
  \v{S}\fi{}irca},\ and\ \citenamefont {Weinriefer}}]{Bernauer}%
  \BibitemOpen
  \bibfield  {author} {\bibinfo {author} {\bibfnamefont {J.~C.}\ \bibnamefont
  {Bernauer}}, \bibinfo {author} {\bibfnamefont {M.~O.}\ \bibnamefont
  {Distler}}, \bibinfo {author} {\bibfnamefont {J.}~\bibnamefont {Friedrich}},
  \bibinfo {author} {\bibfnamefont {T.}~\bibnamefont {Walcher}}, \bibinfo
  {author} {\bibfnamefont {P.}~\bibnamefont {Achenbach}}, \bibinfo {author}
  {\bibfnamefont {C.}~\bibnamefont {Ayerbe~Gayoso}}, \bibinfo {author}
  {\bibfnamefont {R.}~\bibnamefont {B\"ohm}}, \bibinfo {author} {\bibfnamefont
  {D.}~\bibnamefont {Bosnar}}, \bibinfo {author} {\bibfnamefont
  {L.}~\bibnamefont {Debenjak}}, \bibinfo {author} {\bibfnamefont
  {L.}~\bibnamefont {Doria}}, \bibinfo {author} {\bibfnamefont
  {A.}~\bibnamefont {Esser}}, \bibinfo {author} {\bibfnamefont
  {H.}~\bibnamefont {Fonvieille}}, \bibinfo {author} {\bibfnamefont
  {M.}~\bibnamefont {GomezRodriguezdelaPaz}}, \bibinfo {author} {\bibfnamefont
  {J.~M.}\ \bibnamefont {Friedrich}}, \bibinfo {author} {\bibfnamefont
  {M.}~\bibnamefont {Makek}}, \bibinfo {author} {\bibfnamefont
  {H.}~\bibnamefont {Merkel}}, \bibinfo {author} {\bibfnamefont {D.~G.}\
  \bibnamefont {Middleton}}, \bibinfo {author} {\bibfnamefont {U.}~\bibnamefont
  {M\"uller}}, \bibinfo {author} {\bibfnamefont {L.}~\bibnamefont {Nungesser}},
  \bibinfo {author} {\bibfnamefont {J.}~\bibnamefont {Pochodzalla}}, \bibinfo
  {author} {\bibfnamefont {M.}~\bibnamefont {Potokar}}, \bibinfo {author}
  {\bibfnamefont {S.}~\bibnamefont {S\'anchez~Majos}}, \bibinfo {author}
  {\bibfnamefont {B.~S.}\ \bibnamefont {Schlimme}}, \bibinfo {author}
  {\bibfnamefont {S.}~\bibnamefont {\ifmmode~\check{S}\else \v{S}\fi{}irca}}, \
  and\ \bibinfo {author} {\bibfnamefont {M.}~\bibnamefont {Weinriefer}}
  (\bibinfo {collaboration} {A1 Collaboration}),\ }\href {\doibase
  10.1103/PhysRevC.90.015206} {\bibfield  {journal} {\bibinfo  {journal} {Phys.
  Rev. C}\ }\textbf {\bibinfo {volume} {90}},\ \bibinfo {pages} {015206}
  (\bibinfo {year} {2014})}\BibitemShut {NoStop}%
\bibitem [{\citenamefont {Beyer}\ and\ \citenamefont {et~al.}(2017)}]{H-exp}%
  \BibitemOpen
  \bibfield  {author} {\bibinfo {author} {\bibfnamefont {A.}~\bibnamefont
  {Beyer}}\ and\ \bibinfo {author} {\bibnamefont {et~al.}},\ }\href {\doibase
  10.1126/science.aah6677} {\bibfield  {journal} {\bibinfo  {journal}
  {Science}\ }\textbf {\bibinfo {volume} {358}},\ \bibinfo {pages} {79}
  (\bibinfo {year} {2017})}\BibitemShut {NoStop}%
\bibitem [{\citenamefont {Bezginov}\ \emph {et~al.}(2019)\citenamefont
  {Bezginov}, \citenamefont {Valdez}, \citenamefont {Horbatsch}, \citenamefont
  {Marsman}, \citenamefont {Vutha},\ and\ \citenamefont
  {Hessels}}]{LambShift-H}%
  \BibitemOpen
  \bibfield  {author} {\bibinfo {author} {\bibfnamefont {N.}~\bibnamefont
  {Bezginov}}, \bibinfo {author} {\bibfnamefont {T.}~\bibnamefont {Valdez}},
  \bibinfo {author} {\bibfnamefont {M.}~\bibnamefont {Horbatsch}}, \bibinfo
  {author} {\bibfnamefont {A.}~\bibnamefont {Marsman}}, \bibinfo {author}
  {\bibfnamefont {A.~C.}\ \bibnamefont {Vutha}}, \ and\ \bibinfo {author}
  {\bibfnamefont {E.~A.}\ \bibnamefont {Hessels}},\ }\href {\doibase
  10.1126/science.aau7807} {\bibfield  {journal} {\bibinfo  {journal}
  {Science}\ }\textbf {\bibinfo {volume} {365}},\ \bibinfo {pages} {1007}
  (\bibinfo {year} {2019})},\ \Eprint
  {http://arxiv.org/abs/https://science.sciencemag.org/content/365/6457/1007.full.pdf}
  {https://science.sciencemag.org/content/365/6457/1007.full.pdf} \BibitemShut
  {NoStop}%
\bibitem [{\citenamefont {Xiong}\ \emph {et~al.}(2019)\citenamefont {Xiong},
  \citenamefont {Gasparian}, \citenamefont {Gao}, \citenamefont {Dutta},
  \citenamefont {Khandaker}, \citenamefont {Liyanage}, \citenamefont {Pasyuk},
  \citenamefont {Peng}, \citenamefont {Bai}, \citenamefont {Ye}, \citenamefont
  {Gnanvo}, \citenamefont {Gu}, \citenamefont {Levillain}, \citenamefont {Yan},
  \citenamefont {Higinbotham}, \citenamefont {Meziane}, \citenamefont {Ye},
  \citenamefont {Adhikari}, \citenamefont {Aljawrneh}, \citenamefont {Bhatt},
  \citenamefont {Bhetuwal}, \citenamefont {Brock}, \citenamefont {Burkert},
  \citenamefont {Carlin}, \citenamefont {Deur}, \citenamefont {Di},
  \citenamefont {Dunne}, \citenamefont {Ekanayaka}, \citenamefont {El-Fassi},
  \citenamefont {Emmich}, \citenamefont {Gan}, \citenamefont {Glamazdin},
  \citenamefont {Kabir}, \citenamefont {Karki}, \citenamefont {Keith},
  \citenamefont {Kowalski}, \citenamefont {Lagerquist}, \citenamefont {Larin},
  \citenamefont {Liu}, \citenamefont {Liyanage}, \citenamefont {Maxwell},
  \citenamefont {Meekins}, \citenamefont {Nazeer}, \citenamefont {Nelyubin},
  \citenamefont {Nguyen}, \citenamefont {Pedroni}, \citenamefont {Perdrisat},
  \citenamefont {Pierce}, \citenamefont {Punjabi}, \citenamefont {Shabestari},
  \citenamefont {Shahinyan}, \citenamefont {Silwal}, \citenamefont {Stepanyan},
  \citenamefont {Subedi}, \citenamefont {Tarasov}, \citenamefont {Ton},
  \citenamefont {Zhang},\ and\ \citenamefont {Zhao}}]{e-p_scatt}%
  \BibitemOpen
  \bibfield  {author} {\bibinfo {author} {\bibfnamefont {W.}~\bibnamefont
  {Xiong}}, \bibinfo {author} {\bibfnamefont {A.}~\bibnamefont {Gasparian}},
  \bibinfo {author} {\bibfnamefont {H.}~\bibnamefont {Gao}}, \bibinfo {author}
  {\bibfnamefont {D.}~\bibnamefont {Dutta}}, \bibinfo {author} {\bibfnamefont
  {M.}~\bibnamefont {Khandaker}}, \bibinfo {author} {\bibfnamefont
  {N.}~\bibnamefont {Liyanage}}, \bibinfo {author} {\bibfnamefont
  {E.}~\bibnamefont {Pasyuk}}, \bibinfo {author} {\bibfnamefont
  {C.}~\bibnamefont {Peng}}, \bibinfo {author} {\bibfnamefont {X.}~\bibnamefont
  {Bai}}, \bibinfo {author} {\bibfnamefont {L.}~\bibnamefont {Ye}}, \bibinfo
  {author} {\bibfnamefont {K.}~\bibnamefont {Gnanvo}}, \bibinfo {author}
  {\bibfnamefont {C.}~\bibnamefont {Gu}}, \bibinfo {author} {\bibfnamefont
  {M.}~\bibnamefont {Levillain}}, \bibinfo {author} {\bibfnamefont
  {X.}~\bibnamefont {Yan}}, \bibinfo {author} {\bibfnamefont {D.~W.}\
  \bibnamefont {Higinbotham}}, \bibinfo {author} {\bibfnamefont
  {M.}~\bibnamefont {Meziane}}, \bibinfo {author} {\bibfnamefont
  {Z.}~\bibnamefont {Ye}}, \bibinfo {author} {\bibfnamefont {K.}~\bibnamefont
  {Adhikari}}, \bibinfo {author} {\bibfnamefont {B.}~\bibnamefont {Aljawrneh}},
  \bibinfo {author} {\bibfnamefont {H.}~\bibnamefont {Bhatt}}, \bibinfo
  {author} {\bibfnamefont {D.}~\bibnamefont {Bhetuwal}}, \bibinfo {author}
  {\bibfnamefont {J.}~\bibnamefont {Brock}}, \bibinfo {author} {\bibfnamefont
  {V.}~\bibnamefont {Burkert}}, \bibinfo {author} {\bibfnamefont
  {C.}~\bibnamefont {Carlin}}, \bibinfo {author} {\bibfnamefont
  {A.}~\bibnamefont {Deur}}, \bibinfo {author} {\bibfnamefont {D.}~\bibnamefont
  {Di}}, \bibinfo {author} {\bibfnamefont {J.}~\bibnamefont {Dunne}}, \bibinfo
  {author} {\bibfnamefont {P.}~\bibnamefont {Ekanayaka}}, \bibinfo {author}
  {\bibfnamefont {L.}~\bibnamefont {El-Fassi}}, \bibinfo {author}
  {\bibfnamefont {B.}~\bibnamefont {Emmich}}, \bibinfo {author} {\bibfnamefont
  {L.}~\bibnamefont {Gan}}, \bibinfo {author} {\bibfnamefont {O.}~\bibnamefont
  {Glamazdin}}, \bibinfo {author} {\bibfnamefont {M.~L.}\ \bibnamefont
  {Kabir}}, \bibinfo {author} {\bibfnamefont {A.}~\bibnamefont {Karki}},
  \bibinfo {author} {\bibfnamefont {C.}~\bibnamefont {Keith}}, \bibinfo
  {author} {\bibfnamefont {S.}~\bibnamefont {Kowalski}}, \bibinfo {author}
  {\bibfnamefont {V.}~\bibnamefont {Lagerquist}}, \bibinfo {author}
  {\bibfnamefont {I.}~\bibnamefont {Larin}}, \bibinfo {author} {\bibfnamefont
  {T.}~\bibnamefont {Liu}}, \bibinfo {author} {\bibfnamefont {A.}~\bibnamefont
  {Liyanage}}, \bibinfo {author} {\bibfnamefont {J.}~\bibnamefont {Maxwell}},
  \bibinfo {author} {\bibfnamefont {D.}~\bibnamefont {Meekins}}, \bibinfo
  {author} {\bibfnamefont {S.~J.}\ \bibnamefont {Nazeer}}, \bibinfo {author}
  {\bibfnamefont {V.}~\bibnamefont {Nelyubin}}, \bibinfo {author}
  {\bibfnamefont {H.}~\bibnamefont {Nguyen}}, \bibinfo {author} {\bibfnamefont
  {R.}~\bibnamefont {Pedroni}}, \bibinfo {author} {\bibfnamefont
  {C.}~\bibnamefont {Perdrisat}}, \bibinfo {author} {\bibfnamefont
  {J.}~\bibnamefont {Pierce}}, \bibinfo {author} {\bibfnamefont
  {V.}~\bibnamefont {Punjabi}}, \bibinfo {author} {\bibfnamefont
  {M.}~\bibnamefont {Shabestari}}, \bibinfo {author} {\bibfnamefont
  {A.}~\bibnamefont {Shahinyan}}, \bibinfo {author} {\bibfnamefont
  {R.}~\bibnamefont {Silwal}}, \bibinfo {author} {\bibfnamefont
  {S.}~\bibnamefont {Stepanyan}}, \bibinfo {author} {\bibfnamefont
  {A.}~\bibnamefont {Subedi}}, \bibinfo {author} {\bibfnamefont {V.~V.}\
  \bibnamefont {Tarasov}}, \bibinfo {author} {\bibfnamefont {N.}~\bibnamefont
  {Ton}}, \bibinfo {author} {\bibfnamefont {Y.}~\bibnamefont {Zhang}}, \ and\
  \bibinfo {author} {\bibfnamefont {Z.~W.}\ \bibnamefont {Zhao}},\ }\href
  {\doibase 10.1038/s41586-019-1721-2} {\bibfield  {journal} {\bibinfo
  {journal} {Nature (London)}\ }\textbf {\bibinfo {volume} {575}} (\bibinfo
  {year} {2019}),\ 10.1038/s41586-019-1721-2}\BibitemShut {NoStop}%
\bibitem [{\citenamefont {Matveev}\ \emph {et~al.}(2013)\citenamefont
  {Matveev}, \citenamefont {Parthey}, \citenamefont {Predehl}, \citenamefont
  {Alnis}, \citenamefont {Beyer}, \citenamefont {Holzwarth}, \citenamefont
  {Udem}, \citenamefont {Wilken}, \citenamefont {Kolachevsky}, \citenamefont
  {Abgrall}, \citenamefont {Rovera}, \citenamefont {Salomon}, \citenamefont
  {Laurent}, \citenamefont {Grosche}, \citenamefont {Terra}, \citenamefont
  {Legero}, \citenamefont {Schnatz}, \citenamefont {Weyers}, \citenamefont
  {Altschul},\ and\ \citenamefont {H\"ansch}}]{Mat}%
  \BibitemOpen
  \bibfield  {author} {\bibinfo {author} {\bibfnamefont {A.}~\bibnamefont
  {Matveev}}, \bibinfo {author} {\bibfnamefont {C.~G.}\ \bibnamefont
  {Parthey}}, \bibinfo {author} {\bibfnamefont {K.}~\bibnamefont {Predehl}},
  \bibinfo {author} {\bibfnamefont {J.}~\bibnamefont {Alnis}}, \bibinfo
  {author} {\bibfnamefont {A.}~\bibnamefont {Beyer}}, \bibinfo {author}
  {\bibfnamefont {R.}~\bibnamefont {Holzwarth}}, \bibinfo {author}
  {\bibfnamefont {T.}~\bibnamefont {Udem}}, \bibinfo {author} {\bibfnamefont
  {T.}~\bibnamefont {Wilken}}, \bibinfo {author} {\bibfnamefont
  {N.}~\bibnamefont {Kolachevsky}}, \bibinfo {author} {\bibfnamefont
  {M.}~\bibnamefont {Abgrall}}, \bibinfo {author} {\bibfnamefont
  {D.}~\bibnamefont {Rovera}}, \bibinfo {author} {\bibfnamefont
  {C.}~\bibnamefont {Salomon}}, \bibinfo {author} {\bibfnamefont
  {P.}~\bibnamefont {Laurent}}, \bibinfo {author} {\bibfnamefont
  {G.}~\bibnamefont {Grosche}}, \bibinfo {author} {\bibfnamefont
  {O.}~\bibnamefont {Terra}}, \bibinfo {author} {\bibfnamefont
  {T.}~\bibnamefont {Legero}}, \bibinfo {author} {\bibfnamefont
  {H.}~\bibnamefont {Schnatz}}, \bibinfo {author} {\bibfnamefont
  {S.}~\bibnamefont {Weyers}}, \bibinfo {author} {\bibfnamefont
  {B.}~\bibnamefont {Altschul}}, \ and\ \bibinfo {author} {\bibfnamefont
  {T.~W.}\ \bibnamefont {H\"ansch}},\ }\href {\doibase
  10.1103/PhysRevLett.110.230801} {\bibfield  {journal} {\bibinfo  {journal}
  {Phys. Rev. Lett.}\ }\textbf {\bibinfo {volume} {110}},\ \bibinfo {pages}
  {230801} (\bibinfo {year} {2013})}\BibitemShut {NoStop}%
\bibitem [{\citenamefont {Arnoult}\ \emph {et~al.}(2010)\citenamefont
  {Arnoult}, \citenamefont {Nez}, \citenamefont {Julien},\ and\ \citenamefont
  {Biraben}}]{arnoult-1s-3s}%
  \BibitemOpen
  \bibfield  {author} {\bibinfo {author} {\bibfnamefont {O.}~\bibnamefont
  {Arnoult}}, \bibinfo {author} {\bibfnamefont {F.}~\bibnamefont {Nez}},
  \bibinfo {author} {\bibfnamefont {L.}~\bibnamefont {Julien}}, \ and\ \bibinfo
  {author} {\bibfnamefont {F.}~\bibnamefont {Biraben}},\ }\href {\doibase
  10.1140/epjd/e2010-00249-6} {\bibfield  {journal} {\bibinfo  {journal} {{The
  European Physical Journal D : Atomic, molecular, optical and plasma
  physics}}\ }\textbf {\bibinfo {volume} {60}},\ \bibinfo {pages} {243}
  (\bibinfo {year} {2010})}\BibitemShut {NoStop}%
\bibitem [{\citenamefont {Fleurbaey}\ \emph {et~al.}(2018)\citenamefont
  {Fleurbaey}, \citenamefont {Galtier}, \citenamefont {Thomas}, \citenamefont
  {Bonnaud}, \citenamefont {Julien}, \citenamefont {Biraben}, \citenamefont
  {Nez}, \citenamefont {Abgrall},\ and\ \citenamefont
  {Gu\'ena}}]{Fleurbaey-1s-3s}%
  \BibitemOpen
  \bibfield  {author} {\bibinfo {author} {\bibfnamefont {H.}~\bibnamefont
  {Fleurbaey}}, \bibinfo {author} {\bibfnamefont {S.}~\bibnamefont {Galtier}},
  \bibinfo {author} {\bibfnamefont {S.}~\bibnamefont {Thomas}}, \bibinfo
  {author} {\bibfnamefont {M.}~\bibnamefont {Bonnaud}}, \bibinfo {author}
  {\bibfnamefont {L.}~\bibnamefont {Julien}}, \bibinfo {author} {\bibfnamefont
  {F.}~\bibnamefont {Biraben}}, \bibinfo {author} {\bibfnamefont
  {F.}~\bibnamefont {Nez}}, \bibinfo {author} {\bibfnamefont {M.}~\bibnamefont
  {Abgrall}}, \ and\ \bibinfo {author} {\bibfnamefont {J.}~\bibnamefont
  {Gu\'ena}},\ }\href {\doibase 10.1103/PhysRevLett.120.183001} {\bibfield
  {journal} {\bibinfo  {journal} {Phys. Rev. Lett.}\ }\textbf {\bibinfo
  {volume} {120}},\ \bibinfo {pages} {183001} (\bibinfo {year}
  {2018})}\BibitemShut {NoStop}%
\bibitem [{\citenamefont {Low}(1952)}]{Low}%
  \BibitemOpen
  \bibfield  {author} {\bibinfo {author} {\bibfnamefont {F.}~\bibnamefont
  {Low}},\ }\href {\doibase 10.1103/PhysRev.88.53} {\bibfield  {journal}
  {\bibinfo  {journal} {Phys. Rev.}\ }\textbf {\bibinfo {volume} {88}},\
  \bibinfo {pages} {53} (\bibinfo {year} {1952})}\BibitemShut {NoStop}%
\bibitem [{\citenamefont {Labzowsky}\ \emph {et~al.}(1994)\citenamefont
  {Labzowsky}, \citenamefont {Karasiev},\ and\ \citenamefont {Goidenko}}]{LKG}%
  \BibitemOpen
  \bibfield  {author} {\bibinfo {author} {\bibfnamefont {L.}~\bibnamefont
  {Labzowsky}}, \bibinfo {author} {\bibfnamefont {V.}~\bibnamefont {Karasiev}},
  \ and\ \bibinfo {author} {\bibfnamefont {I.}~\bibnamefont {Goidenko}},\
  }\href {\doibase https://doi.org/10.1088/0953-4075/27/15/001} {\bibfield
  {journal} {\bibinfo  {journal} {J. Phys. B: At. Mol. Opt. Phys.}\ }\textbf
  {\bibinfo {volume} {27}},\ \bibinfo {pages} {L439} (\bibinfo {year}
  {1994})}\BibitemShut {NoStop}%
\bibitem [{\citenamefont {Labzowsky}\ \emph {et~al.}(1997)\citenamefont
  {Labzowsky}, \citenamefont {Goidenko},\ and\ \citenamefont {Liesen}}]{LGL}%
  \BibitemOpen
  \bibfield  {author} {\bibinfo {author} {\bibfnamefont {L.~N.}\ \bibnamefont
  {Labzowsky}}, \bibinfo {author} {\bibfnamefont {I.~A.}\ \bibnamefont
  {Goidenko}}, \ and\ \bibinfo {author} {\bibfnamefont {D.}~\bibnamefont
  {Liesen}},\ }\href {\doibase 10.1088/0031-8949/56/3/007} {\bibfield
  {journal} {\bibinfo  {journal} {Physica Scripta}\ }\textbf {\bibinfo {volume}
  {56}},\ \bibinfo {pages} {271} (\bibinfo {year} {1997})}\BibitemShut
  {NoStop}%
\bibitem [{\citenamefont {Labzowsky}\ \emph {et~al.}(2001)\citenamefont
  {Labzowsky}, \citenamefont {Solovyev}, \citenamefont {Plunien},\ and\
  \citenamefont {Soff}}]{LSPS}%
  \BibitemOpen
  \bibfield  {author} {\bibinfo {author} {\bibfnamefont {L.~N.}\ \bibnamefont
  {Labzowsky}}, \bibinfo {author} {\bibfnamefont {D.~A.}\ \bibnamefont
  {Solovyev}}, \bibinfo {author} {\bibfnamefont {G.}~\bibnamefont {Plunien}}, \
  and\ \bibinfo {author} {\bibfnamefont {G.}~\bibnamefont {Soff}},\ }\href
  {\doibase 10.1103/PhysRevLett.87.143003} {\bibfield  {journal} {\bibinfo
  {journal} {Phys. Rev. Lett.}\ }\textbf {\bibinfo {volume} {87}},\ \bibinfo
  {pages} {143003} (\bibinfo {year} {2001})}\BibitemShut {NoStop}%
\bibitem [{\citenamefont {Labzowsky}\ \emph
  {et~al.}(2002{\natexlab{a}})\citenamefont {Labzowsky}, \citenamefont
  {Solovyev}, \citenamefont {Plunien},\ and\ \citenamefont {Soff}}]{can-LSPS}%
  \BibitemOpen
  \bibfield  {author} {\bibinfo {author} {\bibfnamefont {L.}~\bibnamefont
  {Labzowsky}}, \bibinfo {author} {\bibfnamefont {D.}~\bibnamefont {Solovyev}},
  \bibinfo {author} {\bibfnamefont {G.}~\bibnamefont {Plunien}}, \ and\
  \bibinfo {author} {\bibfnamefont {G.}~\bibnamefont {Soff}},\ }\href@noop {}
  {\bibfield  {journal} {\bibinfo  {journal} {Canadian Journal of Physics}\
  }\textbf {\bibinfo {volume} {80}},\ \bibinfo {pages} {1187} (\bibinfo {year}
  {2002}{\natexlab{a}})}\BibitemShut {NoStop}%
\bibitem [{\citenamefont {Labzowsky}\ \emph
  {et~al.}(2002{\natexlab{b}})\citenamefont {Labzowsky}, \citenamefont
  {Soloviev}, \citenamefont {Plunien},\ and\ \citenamefont {Soff}}]{PRA-LSPS}%
  \BibitemOpen
  \bibfield  {author} {\bibinfo {author} {\bibfnamefont {L.}~\bibnamefont
  {Labzowsky}}, \bibinfo {author} {\bibfnamefont {D.}~\bibnamefont {Soloviev}},
  \bibinfo {author} {\bibfnamefont {G.}~\bibnamefont {Plunien}}, \ and\
  \bibinfo {author} {\bibfnamefont {G.}~\bibnamefont {Soff}},\ }\href {\doibase
  10.1103/PhysRevA.65.054502} {\bibfield  {journal} {\bibinfo  {journal} {Phys.
  Rev. A}\ }\textbf {\bibinfo {volume} {65}},\ \bibinfo {pages} {054502}
  (\bibinfo {year} {2002}{\natexlab{b}})}\BibitemShut {NoStop}%
\bibitem [{\citenamefont {Labzowsky}\ \emph {et~al.}(2007)\citenamefont
  {Labzowsky}, \citenamefont {Schedrin}, \citenamefont {Solovyev},\ and\
  \citenamefont {Plunien}}]{PRL-LSSP}%
  \BibitemOpen
  \bibfield  {author} {\bibinfo {author} {\bibfnamefont {L.}~\bibnamefont
  {Labzowsky}}, \bibinfo {author} {\bibfnamefont {G.}~\bibnamefont {Schedrin}},
  \bibinfo {author} {\bibfnamefont {D.}~\bibnamefont {Solovyev}}, \ and\
  \bibinfo {author} {\bibfnamefont {G.}~\bibnamefont {Plunien}},\ }\href
  {\doibase 10.1103/PhysRevLett.98.203003} {\bibfield  {journal} {\bibinfo
  {journal} {Physical Review Letters}\ }\textbf {\bibinfo {volume} {98}},\
  \bibinfo {pages} {2030032} (\bibinfo {year} {2007})}\BibitemShut {NoStop}%
\bibitem [{\citenamefont {Labzowsky}\ \emph {et~al.}(2009)\citenamefont
  {Labzowsky}, \citenamefont {Schedrin}, \citenamefont {Solovyev},
  \citenamefont {Chernovskaya}, \citenamefont {Plunien},\ and\ \citenamefont
  {Karshenboim}}]{LSSCK-2009}%
  \BibitemOpen
  \bibfield  {author} {\bibinfo {author} {\bibfnamefont {L.}~\bibnamefont
  {Labzowsky}}, \bibinfo {author} {\bibfnamefont {G.}~\bibnamefont {Schedrin}},
  \bibinfo {author} {\bibfnamefont {D.}~\bibnamefont {Solovyev}}, \bibinfo
  {author} {\bibfnamefont {E.}~\bibnamefont {Chernovskaya}}, \bibinfo {author}
  {\bibfnamefont {G.}~\bibnamefont {Plunien}}, \ and\ \bibinfo {author}
  {\bibfnamefont {S.}~\bibnamefont {Karshenboim}},\ }\href {\doibase
  10.1103/PhysRevA.79.052506} {\bibfield  {journal} {\bibinfo  {journal} {Phys.
  Rev. A}\ }\textbf {\bibinfo {volume} {79}},\ \bibinfo {pages} {052506}
  (\bibinfo {year} {2009})}\BibitemShut {NoStop}%
\bibitem [{\citenamefont {Jentschura}\ and\ \citenamefont
  {Mohr}(2002)}]{Jent-Mohr}%
  \BibitemOpen
  \bibfield  {author} {\bibinfo {author} {\bibfnamefont {U.~D.}\ \bibnamefont
  {Jentschura}}\ and\ \bibinfo {author} {\bibfnamefont {P.~J.}\ \bibnamefont
  {Mohr}},\ }\href {\doibase 10.1139/p02-019} {\bibfield  {journal} {\bibinfo
  {journal} {Can. J. of Phys.}\ }\textbf {\bibinfo {volume} {80}},\ \bibinfo
  {pages} {633} (\bibinfo {year} {2002})}\BibitemShut {NoStop}%
\bibitem [{\citenamefont {Horbatsch}\ and\ \citenamefont
  {Hessels}(2010)}]{HH-2010}%
  \BibitemOpen
  \bibfield  {author} {\bibinfo {author} {\bibfnamefont {M.}~\bibnamefont
  {Horbatsch}}\ and\ \bibinfo {author} {\bibfnamefont {E.~A.}\ \bibnamefont
  {Hessels}},\ }\href {\doibase 10.1103/PhysRevA.82.052519} {\bibfield
  {journal} {\bibinfo  {journal} {Phys. Rev. A}\ }\textbf {\bibinfo {volume}
  {82}},\ \bibinfo {pages} {052519} (\bibinfo {year} {2010})}\BibitemShut
  {NoStop}%
\bibitem [{\citenamefont {Horbatsch}\ and\ \citenamefont
  {Hessels}(2011)}]{HH-2011}%
  \BibitemOpen
  \bibfield  {author} {\bibinfo {author} {\bibfnamefont {M.}~\bibnamefont
  {Horbatsch}}\ and\ \bibinfo {author} {\bibfnamefont {E.~A.}\ \bibnamefont
  {Hessels}},\ }\href {\doibase 10.1103/PhysRevA.84.032508} {\bibfield
  {journal} {\bibinfo  {journal} {Phys. Rev. A}\ }\textbf {\bibinfo {volume}
  {84}},\ \bibinfo {pages} {032508} (\bibinfo {year} {2011})}\BibitemShut
  {NoStop}%
\bibitem [{\citenamefont {Sansonetti}\ \emph {et~al.}(2011)\citenamefont
  {Sansonetti}, \citenamefont {Simien}, \citenamefont {Gillaspy}, \citenamefont
  {Tan}, \citenamefont {Brewer}, \citenamefont {Brown}, \citenamefont {Wu},\
  and\ \citenamefont {Porto}}]{Sansonetti}%
  \BibitemOpen
  \bibfield  {author} {\bibinfo {author} {\bibfnamefont {C.~J.}\ \bibnamefont
  {Sansonetti}}, \bibinfo {author} {\bibfnamefont {C.~E.}\ \bibnamefont
  {Simien}}, \bibinfo {author} {\bibfnamefont {J.~D.}\ \bibnamefont
  {Gillaspy}}, \bibinfo {author} {\bibfnamefont {J.~N.}\ \bibnamefont {Tan}},
  \bibinfo {author} {\bibfnamefont {S.~M.}\ \bibnamefont {Brewer}}, \bibinfo
  {author} {\bibfnamefont {R.~C.}\ \bibnamefont {Brown}}, \bibinfo {author}
  {\bibfnamefont {S.}~\bibnamefont {Wu}}, \ and\ \bibinfo {author}
  {\bibfnamefont {J.~V.}\ \bibnamefont {Porto}},\ }\href {\doibase
  10.1103/PhysRevLett.107.023001} {\bibfield  {journal} {\bibinfo  {journal}
  {Phys. Rev. Lett.}\ }\textbf {\bibinfo {volume} {107}},\ \bibinfo {pages}
  {023001} (\bibinfo {year} {2011})}\BibitemShut {NoStop}%
\bibitem [{\citenamefont {Brown}\ \emph {et~al.}(2013)\citenamefont {Brown},
  \citenamefont {Wu}, \citenamefont {Porto}, \citenamefont {Sansonetti},
  \citenamefont {Simien}, \citenamefont {Brewer}, \citenamefont {Tan},\ and\
  \citenamefont {Gillaspy}}]{Brown-2013}%
  \BibitemOpen
  \bibfield  {author} {\bibinfo {author} {\bibfnamefont {R.~C.}\ \bibnamefont
  {Brown}}, \bibinfo {author} {\bibfnamefont {S.}~\bibnamefont {Wu}}, \bibinfo
  {author} {\bibfnamefont {J.~V.}\ \bibnamefont {Porto}}, \bibinfo {author}
  {\bibfnamefont {C.~J.}\ \bibnamefont {Sansonetti}}, \bibinfo {author}
  {\bibfnamefont {C.~E.}\ \bibnamefont {Simien}}, \bibinfo {author}
  {\bibfnamefont {S.~M.}\ \bibnamefont {Brewer}}, \bibinfo {author}
  {\bibfnamefont {J.~N.}\ \bibnamefont {Tan}}, \ and\ \bibinfo {author}
  {\bibfnamefont {J.~D.}\ \bibnamefont {Gillaspy}},\ }\href {\doibase
  10.1103/PhysRevA.87.032504} {\bibfield  {journal} {\bibinfo  {journal} {Phys.
  Rev. A}\ }\textbf {\bibinfo {volume} {87}},\ \bibinfo {pages} {032504}
  (\bibinfo {year} {2013})}\BibitemShut {NoStop}%
\bibitem [{\citenamefont {Marsman}\ \emph {et~al.}(2015)\citenamefont
  {Marsman}, \citenamefont {Horbatsch},\ and\ \citenamefont
  {Hessels}}]{MHH-2015}%
  \BibitemOpen
  \bibfield  {author} {\bibinfo {author} {\bibfnamefont {A.}~\bibnamefont
  {Marsman}}, \bibinfo {author} {\bibfnamefont {M.}~\bibnamefont {Horbatsch}},
  \ and\ \bibinfo {author} {\bibfnamefont {E.~A.}\ \bibnamefont {Hessels}},\
  }\href {\doibase 10.1063/1.4922796} {\bibfield  {journal} {\bibinfo
  {journal} {J. Phys. Chem. Ref. Data}\ }\textbf {\bibinfo {volume} {44}},\
  \bibinfo {pages} {031207} (\bibinfo {year} {2015})}\BibitemShut {NoStop}%
\bibitem [{\citenamefont {Amaro}\ \emph
  {et~al.}(2015{\natexlab{a}})\citenamefont {Amaro}, \citenamefont {Fratini},
  \citenamefont {Safari}, \citenamefont {Antognini}, \citenamefont
  {Indelicato}, \citenamefont {Pohl},\ and\ \citenamefont
  {Santos}}]{Amaro-2015}%
  \BibitemOpen
  \bibfield  {author} {\bibinfo {author} {\bibfnamefont {P.}~\bibnamefont
  {Amaro}}, \bibinfo {author} {\bibfnamefont {F.}~\bibnamefont {Fratini}},
  \bibinfo {author} {\bibfnamefont {L.}~\bibnamefont {Safari}}, \bibinfo
  {author} {\bibfnamefont {A.}~\bibnamefont {Antognini}}, \bibinfo {author}
  {\bibfnamefont {P.}~\bibnamefont {Indelicato}}, \bibinfo {author}
  {\bibfnamefont {R.}~\bibnamefont {Pohl}}, \ and\ \bibinfo {author}
  {\bibfnamefont {J.~P.}\ \bibnamefont {Santos}},\ }\href {\doibase
  10.1103/PhysRevA.92.062506} {\bibfield  {journal} {\bibinfo  {journal} {Phys.
  Rev. A}\ }\textbf {\bibinfo {volume} {92}},\ \bibinfo {pages} {062506}
  (\bibinfo {year} {2015}{\natexlab{a}})}\BibitemShut {NoStop}%
\bibitem [{\citenamefont {Amaro}\ \emph
  {et~al.}(2015{\natexlab{b}})\citenamefont {Amaro}, \citenamefont {Franke},
  \citenamefont {Krauth}, \citenamefont {Diepold}, \citenamefont {Fratini},
  \citenamefont {Safari}, \citenamefont {Machado}, \citenamefont {Antognini},
  \citenamefont {Kottmann}, \citenamefont {Indelicato}, \citenamefont {Pohl},\
  and\ \citenamefont {Santos}}]{Amaro-mH-2015}%
  \BibitemOpen
  \bibfield  {author} {\bibinfo {author} {\bibfnamefont {P.}~\bibnamefont
  {Amaro}}, \bibinfo {author} {\bibfnamefont {B.}~\bibnamefont {Franke}},
  \bibinfo {author} {\bibfnamefont {J.~J.}\ \bibnamefont {Krauth}}, \bibinfo
  {author} {\bibfnamefont {M.}~\bibnamefont {Diepold}}, \bibinfo {author}
  {\bibfnamefont {F.}~\bibnamefont {Fratini}}, \bibinfo {author} {\bibfnamefont
  {L.}~\bibnamefont {Safari}}, \bibinfo {author} {\bibfnamefont
  {J.}~\bibnamefont {Machado}}, \bibinfo {author} {\bibfnamefont
  {A.}~\bibnamefont {Antognini}}, \bibinfo {author} {\bibfnamefont
  {F.}~\bibnamefont {Kottmann}}, \bibinfo {author} {\bibfnamefont
  {P.}~\bibnamefont {Indelicato}}, \bibinfo {author} {\bibfnamefont
  {R.}~\bibnamefont {Pohl}}, \ and\ \bibinfo {author} {\bibfnamefont {J.~P.}\
  \bibnamefont {Santos}},\ }\href {\doibase 10.1103/PhysRevA.92.022514}
  {\bibfield  {journal} {\bibinfo  {journal} {Phys. Rev. A}\ }\textbf {\bibinfo
  {volume} {92}},\ \bibinfo {pages} {022514} (\bibinfo {year}
  {2015}{\natexlab{b}})}\BibitemShut {NoStop}%
\bibitem [{\citenamefont {Solovyev}\ \emph {et~al.}(2020)\citenamefont
  {Solovyev}, \citenamefont {Anikin}, \citenamefont {Zalialiutdinov},\ and\
  \citenamefont {Labzowsky}}]{S-2020-importance}%
  \BibitemOpen
  \bibfield  {author} {\bibinfo {author} {\bibfnamefont {D.}~\bibnamefont
  {Solovyev}}, \bibinfo {author} {\bibfnamefont {A.}~\bibnamefont {Anikin}},
  \bibinfo {author} {\bibfnamefont {T.}~\bibnamefont {Zalialiutdinov}}, \ and\
  \bibinfo {author} {\bibfnamefont {L.}~\bibnamefont {Labzowsky}},\ }\href
  {\doibase 10.1088/1361-6455/ab8b43} {\bibfield  {journal} {\bibinfo
  {journal} {Journal of Physics B: Atomic, Molecular and Optical Physics}\
  }\textbf {\bibinfo {volume} {53}},\ \bibinfo {pages} {125002} (\bibinfo
  {year} {2020})}\BibitemShut {NoStop}%
\bibitem [{\citenamefont {Antognini}\ \emph
  {et~al.}(2013{\natexlab{a}})\citenamefont {Antognini}, \citenamefont {Nez},
  \citenamefont {Schuhmann}, \citenamefont {Amaro}, \citenamefont {Biraben},
  \citenamefont {Cardoso}, \citenamefont {Covita}, \citenamefont {Dax},
  \citenamefont {Dhawan}, \citenamefont {Diepold}, \citenamefont {Fernandes},
  \citenamefont {Giesen}, \citenamefont {Gouvea}, \citenamefont {Graf},
  \citenamefont {H{\"a}nsch}, \citenamefont {Indelicato}, \citenamefont
  {Julien}, \citenamefont {Kao}, \citenamefont {Knowles}, \citenamefont
  {Kottmann}, \citenamefont {Le~Bigot}, \citenamefont {Liu}, \citenamefont
  {Lopes}, \citenamefont {Ludhova}, \citenamefont {Monteiro}, \citenamefont
  {Mulhauser}, \citenamefont {Nebel}, \citenamefont {Rabinowitz}, \citenamefont
  {dos Santos}, \citenamefont {Schaller}, \citenamefont {Schwob}, \citenamefont
  {Taqqu}, \citenamefont {Veloso}, \citenamefont {Vogelsang},\ and\
  \citenamefont {Pohl}}]{Antognini-2013}%
  \BibitemOpen
  \bibfield  {author} {\bibinfo {author} {\bibfnamefont {A.}~\bibnamefont
  {Antognini}}, \bibinfo {author} {\bibfnamefont {F.}~\bibnamefont {Nez}},
  \bibinfo {author} {\bibfnamefont {K.}~\bibnamefont {Schuhmann}}, \bibinfo
  {author} {\bibfnamefont {F.~D.}\ \bibnamefont {Amaro}}, \bibinfo {author}
  {\bibfnamefont {F.}~\bibnamefont {Biraben}}, \bibinfo {author} {\bibfnamefont
  {J.~M.~R.}\ \bibnamefont {Cardoso}}, \bibinfo {author} {\bibfnamefont
  {D.~S.}\ \bibnamefont {Covita}}, \bibinfo {author} {\bibfnamefont
  {A.}~\bibnamefont {Dax}}, \bibinfo {author} {\bibfnamefont {S.}~\bibnamefont
  {Dhawan}}, \bibinfo {author} {\bibfnamefont {M.}~\bibnamefont {Diepold}},
  \bibinfo {author} {\bibfnamefont {L.~M.~P.}\ \bibnamefont {Fernandes}},
  \bibinfo {author} {\bibfnamefont {A.}~\bibnamefont {Giesen}}, \bibinfo
  {author} {\bibfnamefont {A.~L.}\ \bibnamefont {Gouvea}}, \bibinfo {author}
  {\bibfnamefont {T.}~\bibnamefont {Graf}}, \bibinfo {author} {\bibfnamefont
  {T.~W.}\ \bibnamefont {H{\"a}nsch}}, \bibinfo {author} {\bibfnamefont
  {P.}~\bibnamefont {Indelicato}}, \bibinfo {author} {\bibfnamefont
  {L.}~\bibnamefont {Julien}}, \bibinfo {author} {\bibfnamefont {C.-Y.}\
  \bibnamefont {Kao}}, \bibinfo {author} {\bibfnamefont {P.}~\bibnamefont
  {Knowles}}, \bibinfo {author} {\bibfnamefont {F.}~\bibnamefont {Kottmann}},
  \bibinfo {author} {\bibfnamefont {E.-O.}\ \bibnamefont {Le~Bigot}}, \bibinfo
  {author} {\bibfnamefont {Y.-W.}\ \bibnamefont {Liu}}, \bibinfo {author}
  {\bibfnamefont {J.~A.~M.}\ \bibnamefont {Lopes}}, \bibinfo {author}
  {\bibfnamefont {L.}~\bibnamefont {Ludhova}}, \bibinfo {author} {\bibfnamefont
  {C.~M.~B.}\ \bibnamefont {Monteiro}}, \bibinfo {author} {\bibfnamefont
  {F.}~\bibnamefont {Mulhauser}}, \bibinfo {author} {\bibfnamefont
  {T.}~\bibnamefont {Nebel}}, \bibinfo {author} {\bibfnamefont
  {P.}~\bibnamefont {Rabinowitz}}, \bibinfo {author} {\bibfnamefont {J.~M.~F.}\
  \bibnamefont {dos Santos}}, \bibinfo {author} {\bibfnamefont {L.~A.}\
  \bibnamefont {Schaller}}, \bibinfo {author} {\bibfnamefont {C.}~\bibnamefont
  {Schwob}}, \bibinfo {author} {\bibfnamefont {D.}~\bibnamefont {Taqqu}},
  \bibinfo {author} {\bibfnamefont {J.~F. C.~A.}\ \bibnamefont {Veloso}},
  \bibinfo {author} {\bibfnamefont {J.}~\bibnamefont {Vogelsang}}, \ and\
  \bibinfo {author} {\bibfnamefont {R.}~\bibnamefont {Pohl}},\ }\href {\doibase
  10.1126/science.1230016} {\bibfield  {journal} {\bibinfo  {journal}
  {Science}\ }\textbf {\bibinfo {volume} {339}},\ \bibinfo {pages} {417}
  (\bibinfo {year} {2013}{\natexlab{a}})},\ \Eprint
  {http://arxiv.org/abs/https://science.sciencemag.org/content/339/6118/417.full.pdf}
  {https://science.sciencemag.org/content/339/6118/417.full.pdf} \BibitemShut
  {NoStop}%
\bibitem [{\citenamefont {Antognini}\ \emph
  {et~al.}(2013{\natexlab{b}})\citenamefont {Antognini}, \citenamefont
  {Kottmann}, \citenamefont {Biraben}, \citenamefont {Indelicato},
  \citenamefont {Nez},\ and\ \citenamefont {Pohl}}]{Antog-2013}%
  \BibitemOpen
  \bibfield  {author} {\bibinfo {author} {\bibfnamefont {A.}~\bibnamefont
  {Antognini}}, \bibinfo {author} {\bibfnamefont {F.}~\bibnamefont {Kottmann}},
  \bibinfo {author} {\bibfnamefont {F.}~\bibnamefont {Biraben}}, \bibinfo
  {author} {\bibfnamefont {P.}~\bibnamefont {Indelicato}}, \bibinfo {author}
  {\bibfnamefont {F.}~\bibnamefont {Nez}}, \ and\ \bibinfo {author}
  {\bibfnamefont {R.}~\bibnamefont {Pohl}},\ }\href {\doibase
  https://doi.org/10.1016/j.aop.2012.12.003} {\bibfield  {journal} {\bibinfo
  {journal} {Annals of Physics}\ }\textbf {\bibinfo {volume} {331}},\ \bibinfo
  {pages} {127 } (\bibinfo {year} {2013}{\natexlab{b}})}\BibitemShut {NoStop}%
\bibitem [{\citenamefont {Pachucki}(1996)}]{Pachucki}%
  \BibitemOpen
  \bibfield  {author} {\bibinfo {author} {\bibfnamefont {K.}~\bibnamefont
  {Pachucki}},\ }\href {\doibase 10.1103/PhysRevA.53.2092} {\bibfield
  {journal} {\bibinfo  {journal} {Phys. Rev. A}\ }\textbf {\bibinfo {volume}
  {53}},\ \bibinfo {pages} {2092} (\bibinfo {year} {1996})}\BibitemShut
  {NoStop}%
\bibitem [{\citenamefont {Martynenko}(2008)}]{Martynenko-2008}%
  \BibitemOpen
  \bibfield  {author} {\bibinfo {author} {\bibfnamefont {A.}~\bibnamefont
  {Martynenko}},\ }\href {\doibase 10.1007/s11450-008-1014-y} {\bibfield
  {journal} {\bibinfo  {journal} {Phys. Atom. Nucl.}\ }\textbf {\bibinfo
  {volume} {71}},\ \bibinfo {pages} {125} (\bibinfo {year} {2008})},\ \Eprint
  {http://arxiv.org/abs/hep-ph/0610226} {arXiv:hep-ph/0610226} \BibitemShut
  {NoStop}%
\bibitem [{\citenamefont {Karshenboim}\ \emph {et~al.}(2015)\citenamefont
  {Karshenboim}, \citenamefont {Korzinin}, \citenamefont {Shelyuto},\ and\
  \citenamefont {Ivanov}}]{Karshenboim}%
  \BibitemOpen
  \bibfield  {author} {\bibinfo {author} {\bibfnamefont {S.~G.}\ \bibnamefont
  {Karshenboim}}, \bibinfo {author} {\bibfnamefont {E.~Y.}\ \bibnamefont
  {Korzinin}}, \bibinfo {author} {\bibfnamefont {V.~A.}\ \bibnamefont
  {Shelyuto}}, \ and\ \bibinfo {author} {\bibfnamefont {V.~G.}\ \bibnamefont
  {Ivanov}},\ }\href {\doibase 10.1063/1.4921197} {\bibfield  {journal}
  {\bibinfo  {journal} {Journal of Physical and Chemical Reference Data}\
  }\textbf {\bibinfo {volume} {44}},\ \bibinfo {pages} {031202} (\bibinfo
  {year} {2015})},\ \Eprint
  {http://arxiv.org/abs/https://doi.org/10.1063/1.4921197}
  {https://doi.org/10.1063/1.4921197} \BibitemShut {NoStop}%
\bibitem [{\citenamefont {Labzowsky}\ \emph {et~al.}(2003)\citenamefont
  {Labzowsky}, \citenamefont {Solovyev}, \citenamefont {Sharipov},
  \citenamefont {Plunien},\ and\ \citenamefont {Soff}}]{LSSPS-2003}%
  \BibitemOpen
  \bibfield  {author} {\bibinfo {author} {\bibfnamefont {L.}~\bibnamefont
  {Labzowsky}}, \bibinfo {author} {\bibfnamefont {D.}~\bibnamefont {Solovyev}},
  \bibinfo {author} {\bibfnamefont {V.}~\bibnamefont {Sharipov}}, \bibinfo
  {author} {\bibfnamefont {G.}~\bibnamefont {Plunien}}, \ and\ \bibinfo
  {author} {\bibfnamefont {G.}~\bibnamefont {Soff}},\ }\href {\doibase
  10.1088/0953-4075/36/15/101} {\bibfield  {journal} {\bibinfo  {journal} {J.
  Phys. B: At., Mol. and Opt. Phys.}\ }\textbf {\bibinfo {volume} {36}},\
  \bibinfo {pages} {L227} (\bibinfo {year} {2003})}\BibitemShut {NoStop}%
\bibitem [{\citenamefont {Weitz}\ \emph {et~al.}(1995)\citenamefont {Weitz},
  \citenamefont {Huber}, \citenamefont {Schmidt-Kaler}, \citenamefont
  {Leibfried}, \citenamefont {Vassen}, \citenamefont {Zimmermann},
  \citenamefont {Pachucki}, \citenamefont {H\"ansch}, \citenamefont {Julien},\
  and\ \citenamefont {Biraben}}]{Weitz}%
  \BibitemOpen
  \bibfield  {author} {\bibinfo {author} {\bibfnamefont {M.}~\bibnamefont
  {Weitz}}, \bibinfo {author} {\bibfnamefont {A.}~\bibnamefont {Huber}},
  \bibinfo {author} {\bibfnamefont {F.}~\bibnamefont {Schmidt-Kaler}}, \bibinfo
  {author} {\bibfnamefont {D.}~\bibnamefont {Leibfried}}, \bibinfo {author}
  {\bibfnamefont {W.}~\bibnamefont {Vassen}}, \bibinfo {author} {\bibfnamefont
  {C.}~\bibnamefont {Zimmermann}}, \bibinfo {author} {\bibfnamefont
  {K.}~\bibnamefont {Pachucki}}, \bibinfo {author} {\bibfnamefont {T.~W.}\
  \bibnamefont {H\"ansch}}, \bibinfo {author} {\bibfnamefont {L.}~\bibnamefont
  {Julien}}, \ and\ \bibinfo {author} {\bibfnamefont {F.}~\bibnamefont
  {Biraben}},\ }\href {\doibase 10.1103/PhysRevA.52.2664} {\bibfield  {journal}
  {\bibinfo  {journal} {Phys. Rev. A}\ }\textbf {\bibinfo {volume} {52}},\
  \bibinfo {pages} {2664} (\bibinfo {year} {1995})}\BibitemShut {NoStop}%
\bibitem [{\citenamefont {Weitz}\ \emph {et~al.}(1992)\citenamefont {Weitz},
  \citenamefont {Schmidt-Kaler},\ and\ \citenamefont {H\"ansch}}]{Weitz-1}%
  \BibitemOpen
  \bibfield  {author} {\bibinfo {author} {\bibfnamefont {M.}~\bibnamefont
  {Weitz}}, \bibinfo {author} {\bibfnamefont {F.}~\bibnamefont
  {Schmidt-Kaler}}, \ and\ \bibinfo {author} {\bibfnamefont {T.~W.}\
  \bibnamefont {H\"ansch}},\ }\href {\doibase 10.1103/PhysRevLett.68.1120}
  {\bibfield  {journal} {\bibinfo  {journal} {Phys. Rev. Lett.}\ }\textbf
  {\bibinfo {volume} {68}},\ \bibinfo {pages} {1120} (\bibinfo {year}
  {1992})}\BibitemShut {NoStop}%
\bibitem [{\citenamefont {Nez}\ \emph {et~al.}(1992)\citenamefont {Nez},
  \citenamefont {Plimmer}, \citenamefont {Bourzeix}, \citenamefont {Julien},
  \citenamefont {Biraben}, \citenamefont {Felder}, \citenamefont {Acef},
  \citenamefont {Zondy}, \citenamefont {Laurent}, \citenamefont {Clairon},
  \citenamefont {Abed}, \citenamefont {Millerioux},\ and\ \citenamefont
  {Juncar}}]{Nez}%
  \BibitemOpen
  \bibfield  {author} {\bibinfo {author} {\bibfnamefont {F.}~\bibnamefont
  {Nez}}, \bibinfo {author} {\bibfnamefont {M.~D.}\ \bibnamefont {Plimmer}},
  \bibinfo {author} {\bibfnamefont {S.}~\bibnamefont {Bourzeix}}, \bibinfo
  {author} {\bibfnamefont {L.}~\bibnamefont {Julien}}, \bibinfo {author}
  {\bibfnamefont {F.}~\bibnamefont {Biraben}}, \bibinfo {author} {\bibfnamefont
  {R.}~\bibnamefont {Felder}}, \bibinfo {author} {\bibfnamefont
  {O.}~\bibnamefont {Acef}}, \bibinfo {author} {\bibfnamefont {J.~J.}\
  \bibnamefont {Zondy}}, \bibinfo {author} {\bibfnamefont {P.}~\bibnamefont
  {Laurent}}, \bibinfo {author} {\bibfnamefont {A.}~\bibnamefont {Clairon}},
  \bibinfo {author} {\bibfnamefont {M.}~\bibnamefont {Abed}}, \bibinfo {author}
  {\bibfnamefont {Y.}~\bibnamefont {Millerioux}}, \ and\ \bibinfo {author}
  {\bibfnamefont {P.}~\bibnamefont {Juncar}},\ }\href {\doibase
  10.1103/PhysRevLett.69.2326} {\bibfield  {journal} {\bibinfo  {journal}
  {Phys. Rev. Lett.}\ }\textbf {\bibinfo {volume} {69}},\ \bibinfo {pages}
  {2326} (\bibinfo {year} {1992})}\BibitemShut {NoStop}%
\bibitem [{\citenamefont {Anikin}\ \emph {et~al.}(2021)\citenamefont {Anikin},
  \citenamefont {Zalialiutdinov},\ and\ \citenamefont {Solovyev}}]{Anikin}%
  \BibitemOpen
  \bibfield  {author} {\bibinfo {author} {\bibfnamefont {A.}~\bibnamefont
  {Anikin}}, \bibinfo {author} {\bibfnamefont {T.}~\bibnamefont
  {Zalialiutdinov}}, \ and\ \bibinfo {author} {\bibfnamefont {D.}~\bibnamefont
  {Solovyev}},\ }\href {\doibase 10.1103/PhysRevA.103.022833} {\bibfield
  {journal} {\bibinfo  {journal} {Phys. Rev. A}\ }\textbf {\bibinfo {volume}
  {103}},\ \bibinfo {pages} {022833} (\bibinfo {year} {2021})}\BibitemShut
  {NoStop}%
\bibitem [{\citenamefont {de~Beauvoir}\ \emph {et~al.}(1997)\citenamefont
  {de~Beauvoir}, \citenamefont {Nez}, \citenamefont {Julien}, \citenamefont
  {Cagnac}, \citenamefont {Biraben}, \citenamefont {Touahri}, \citenamefont
  {Hilico}, \citenamefont {Acef}, \citenamefont {Clairon},\ and\ \citenamefont
  {Zondy}}]{deB-0}%
  \BibitemOpen
  \bibfield  {author} {\bibinfo {author} {\bibfnamefont {B.}~\bibnamefont
  {de~Beauvoir}}, \bibinfo {author} {\bibfnamefont {F.}~\bibnamefont {Nez}},
  \bibinfo {author} {\bibfnamefont {L.}~\bibnamefont {Julien}}, \bibinfo
  {author} {\bibfnamefont {B.}~\bibnamefont {Cagnac}}, \bibinfo {author}
  {\bibfnamefont {F.}~\bibnamefont {Biraben}}, \bibinfo {author} {\bibfnamefont
  {D.}~\bibnamefont {Touahri}}, \bibinfo {author} {\bibfnamefont
  {L.}~\bibnamefont {Hilico}}, \bibinfo {author} {\bibfnamefont
  {O.}~\bibnamefont {Acef}}, \bibinfo {author} {\bibfnamefont {A.}~\bibnamefont
  {Clairon}}, \ and\ \bibinfo {author} {\bibfnamefont {J.~J.}\ \bibnamefont
  {Zondy}},\ }\href {\doibase 10.1103/PhysRevLett.78.440} {\bibfield  {journal}
  {\bibinfo  {journal} {Phys. Rev. Lett.}\ }\textbf {\bibinfo {volume} {78}},\
  \bibinfo {pages} {440} (\bibinfo {year} {1997})}\BibitemShut {NoStop}%
\bibitem [{\citenamefont {{Schwob}}\ \emph {et~al.}(1999)\citenamefont
  {{Schwob}}, \citenamefont {{Jozefowski}}, \citenamefont {{Acef}},
  \citenamefont {{Hilico}}, \citenamefont {{de Beauvoir}}, \citenamefont
  {{Nez}}, \citenamefont {{Julien}}, \citenamefont {{Clairon}},\ and\
  \citenamefont {{Biraben}}}]{Schwob}%
  \BibitemOpen
  \bibfield  {author} {\bibinfo {author} {\bibfnamefont {C.}~\bibnamefont
  {{Schwob}}}, \bibinfo {author} {\bibfnamefont {L.}~\bibnamefont
  {{Jozefowski}}}, \bibinfo {author} {\bibfnamefont {O.}~\bibnamefont
  {{Acef}}}, \bibinfo {author} {\bibfnamefont {L.}~\bibnamefont {{Hilico}}},
  \bibinfo {author} {\bibfnamefont {B.}~\bibnamefont {{de Beauvoir}}}, \bibinfo
  {author} {\bibfnamefont {F.}~\bibnamefont {{Nez}}}, \bibinfo {author}
  {\bibfnamefont {L.}~\bibnamefont {{Julien}}}, \bibinfo {author}
  {\bibfnamefont {A.}~\bibnamefont {{Clairon}}}, \ and\ \bibinfo {author}
  {\bibfnamefont {F.}~\bibnamefont {{Biraben}}},\ }\href@noop {} {\bibfield
  {journal} {\bibinfo  {journal} {IEEE Transactions on Instrumentation and
  Measurement}\ }\textbf {\bibinfo {volume} {48}},\ \bibinfo {pages} {178}
  (\bibinfo {year} {1999})}\BibitemShut {NoStop}%
\bibitem [{\citenamefont {Schwob}\ \emph {et~al.}(1999)\citenamefont {Schwob},
  \citenamefont {Jozefowski}, \citenamefont {{de Beauvoir}}, \citenamefont
  {Hilico}, \citenamefont {Nez}, \citenamefont {Julien}, \citenamefont
  {Biraben}, \citenamefont {Acef}, \citenamefont {Zondy},\ and\ \citenamefont
  {Clairon}}]{deB-1}%
  \BibitemOpen
  \bibfield  {author} {\bibinfo {author} {\bibfnamefont {C.}~\bibnamefont
  {Schwob}}, \bibinfo {author} {\bibfnamefont {L.}~\bibnamefont {Jozefowski}},
  \bibinfo {author} {\bibfnamefont {B.}~\bibnamefont {{de Beauvoir}}}, \bibinfo
  {author} {\bibfnamefont {L.}~\bibnamefont {Hilico}}, \bibinfo {author}
  {\bibfnamefont {F.}~\bibnamefont {Nez}}, \bibinfo {author} {\bibfnamefont
  {L.}~\bibnamefont {Julien}}, \bibinfo {author} {\bibfnamefont
  {F.}~\bibnamefont {Biraben}}, \bibinfo {author} {\bibfnamefont
  {O.}~\bibnamefont {Acef}}, \bibinfo {author} {\bibfnamefont {J.-J.}\
  \bibnamefont {Zondy}}, \ and\ \bibinfo {author} {\bibfnamefont
  {A.}~\bibnamefont {Clairon}},\ }\href {\doibase 10.1103/PhysRevLett.82.4960}
  {\bibfield  {journal} {\bibinfo  {journal} {Phys. Rev. Lett.}\ }\textbf
  {\bibinfo {volume} {82}},\ \bibinfo {pages} {4960} (\bibinfo {year}
  {1999})}\BibitemShut {NoStop}%
\bibitem [{\citenamefont {{de Beauvoir}}\ \emph {et~al.}(2000)\citenamefont
  {{de Beauvoir}}, \citenamefont {Schwob}, \citenamefont {Acef}, \citenamefont
  {Jozefowski}, \citenamefont {Hilico}, \citenamefont {Nez}, \citenamefont
  {Julien}, \citenamefont {Clairon},\ and\ \citenamefont {Biraben}}]{deB-2}%
  \BibitemOpen
  \bibfield  {author} {\bibinfo {author} {\bibfnamefont {B.}~\bibnamefont {{de
  Beauvoir}}}, \bibinfo {author} {\bibfnamefont {C.}~\bibnamefont {Schwob}},
  \bibinfo {author} {\bibfnamefont {O.}~\bibnamefont {Acef}}, \bibinfo {author}
  {\bibfnamefont {L.}~\bibnamefont {Jozefowski}}, \bibinfo {author}
  {\bibfnamefont {L.}~\bibnamefont {Hilico}}, \bibinfo {author} {\bibfnamefont
  {F.}~\bibnamefont {Nez}}, \bibinfo {author} {\bibfnamefont {L.}~\bibnamefont
  {Julien}}, \bibinfo {author} {\bibfnamefont {A.}~\bibnamefont {Clairon}}, \
  and\ \bibinfo {author} {\bibfnamefont {F.}~\bibnamefont {Biraben}},\ }\href
  {\doibase 10.1007/s100530070043} {\bibfield  {journal} {\bibinfo  {journal}
  {Eur. Phys. J. D}\ }\textbf {\bibinfo {volume} {12}},\ \bibinfo {pages} {61}
  (\bibinfo {year} {2000})}\BibitemShut {NoStop}%
\bibitem [{\citenamefont {Andreev}\ \emph {et~al.}(2008)\citenamefont
  {Andreev}, \citenamefont {Labzowsky}, \citenamefont {Plunien},\ and\
  \citenamefont {Solovyev}}]{Andr}%
  \BibitemOpen
  \bibfield  {author} {\bibinfo {author} {\bibfnamefont {O.~Y.}\ \bibnamefont
  {Andreev}}, \bibinfo {author} {\bibfnamefont {L.~N.}\ \bibnamefont
  {Labzowsky}}, \bibinfo {author} {\bibfnamefont {G.}~\bibnamefont {Plunien}},
  \ and\ \bibinfo {author} {\bibfnamefont {D.~A.}\ \bibnamefont {Solovyev}},\
  }\href {\doibase 10.1016/j.physrep.2007.10.003} {\bibfield  {journal}
  {\bibinfo  {journal} {Phys. Rep.}\ }\textbf {\bibinfo {volume} {455}},\
  \bibinfo {pages} {135} (\bibinfo {year} {2008})}\BibitemShut {NoStop}%
\bibitem [{\citenamefont {Zalialiutdinov}\ \emph {et~al.}(2018)\citenamefont
  {Zalialiutdinov}, \citenamefont {Solovyev}, \citenamefont {Labzowsky},\ and\
  \citenamefont {Plunien}}]{ZSLP-report}%
  \BibitemOpen
  \bibfield  {author} {\bibinfo {author} {\bibfnamefont {T.~A.}\ \bibnamefont
  {Zalialiutdinov}}, \bibinfo {author} {\bibfnamefont {D.~A.}\ \bibnamefont
  {Solovyev}}, \bibinfo {author} {\bibfnamefont {L.~N.}\ \bibnamefont
  {Labzowsky}}, \ and\ \bibinfo {author} {\bibfnamefont {G.}~\bibnamefont
  {Plunien}},\ }\href {\doibase 10.1016/j.physrep.2018.02.003} {\bibfield
  {journal} {\bibinfo  {journal} {Phys. Rep.}\ }\textbf {\bibinfo {volume}
  {737}},\ \bibinfo {pages} {1 } (\bibinfo {year} {2018})}\BibitemShut
  {NoStop}%
\bibitem [{\citenamefont {Horbatsch}\ and\ \citenamefont
  {Hessels}(2016)}]{HH-tab}%
  \BibitemOpen
  \bibfield  {author} {\bibinfo {author} {\bibfnamefont {M.}~\bibnamefont
  {Horbatsch}}\ and\ \bibinfo {author} {\bibfnamefont {E.~A.}\ \bibnamefont
  {Hessels}},\ }\href {\doibase 10.1103/PhysRevA.93.022513} {\bibfield
  {journal} {\bibinfo  {journal} {Phys. Rev. A}\ }\textbf {\bibinfo {volume}
  {93}},\ \bibinfo {pages} {022513} (\bibinfo {year} {2016})}\BibitemShut
  {NoStop}%
\end{thebibliography}%
\end{document}